\documentclass[10pt]{article}
\usepackage{graphicx}
\begin{document}

\title{Nucleon Resonance Structure Studies Via Exclusive $KY$ Electroproduction}

\author{Daniel S. Carman (for the CLAS Collaboration)\\
carman@jlab.org\\
Jefferson Laboratory, 12000 Jefferson Ave., Newport News VA, 23606, USA}

\maketitle

\begin{abstract}
Studying the structure of excited nucleon states employing the electroproduction of exclusive
reactions is an important avenue for exploring the nature of the non-perturbative strong 
interaction. The electrocouplings of $N^*$ states in the mass range below 1.8~GeV have been 
determined from analyses of CLAS $\pi N$, $\eta N$, and $\pi \pi N$ data. This work has made it 
clear that consistent results from independent analyses of several exclusive channels with different 
couplings and non-resonant backgrounds but the same $N^*$ electro-excitation amplitudes, is essential 
to have confidence in the extracted results. In terms of hadronic coupling, many high-lying $N^*$ 
states preferentially decay through the $\pi \pi N$ channel instead of $\pi N$. Data from the $KY$ 
channels will therefore be critical to provide an independent analysis to compare the extracted 
electrocouplings for the high-lying $N^*$ states against those determined from the $\pi N$ and 
$\pi \pi N$ channels. A program to study excited $N^*$ state structure in both non-strange and 
strange exclusive electroproduction channels using CLAS12 will measure differential cross sections 
and polarization observables to be used as input to extract the $\gamma_vNN^*$ electrocoupling 
amplitudes for the most prominent $N^*$ states in the range of invariant energy $W$ up 3~GeV in 
the virtually unexplored domain of momentum transfers $Q^2$ up to 12~GeV$^2$.
\end{abstract}

\section{Introduction}

Intensive spectroscopy of the nucleon excitation spectrum and detailed studies of the structure of 
these excited states has played a pivotal role in the development of our understanding of the 
strong interaction. The concept of quarks that emerged through such studies led to the development 
of the constituent quark model~\cite{isgur,capstick} (CQM) in the 1980s. As a result of intense 
experimental and theoretical effort over the past 30 years, it is now apparent that the structure 
of the states in the nucleon excitation spectrum is much more complex than what can be described in 
terms of models based on constituent quarks alone. At the typical energy and distance scales found 
within the $N^*$ states, the quark-gluon coupling is large. Therefore, we are confronted with the 
fact that quark-gluon confinement, hadron mass generation, and the dynamics that give rise to the 
$N^*$ spectrum, cannot be understood within the framework of perturbative Quantum Chromodynamics (QCD). 
The need to understand QCD in this non-perturbative domain is a fundamental issue in nuclear physics, 
which the study of $N^*$ structure can help to address. Such studies, in fact, represent a necessary 
step toward understanding how QCD in the regime of large quark-gluon couplings generates mass and 
how systems of confined quarks and gluons, i.e. mesons and baryons, are formed.

Studies of low-lying nucleon excited states using electromagnetic probes at four-momentum transfer 
$Q^2 < 5$~GeV$^2$ have revealed that the structure of these states is a complex 
interplay between the internal core of three dressed quarks and an external meson-baryon cloud. $N^*$ 
states of different quantum numbers have significantly different relative contributions from these 
two components, demonstrating distinctly different manifestations of the non-perturbative strong 
interaction in their generation. The relative contribution of the quark core increases with $Q^2$ in 
a gradual transition to a dominance of quark degrees of freedom for $Q^2 > 5$~GeV$^2$. This kinematics 
area still remains almost unexplored in exclusive reactions. Studies of the $Q^2$ evolution of $N^*$ 
structure from low to high $Q^2$ offer access to the strong interaction between dressed quarks in the 
non-perturbative regime that is responsible for $N^*$ formation.

Electroproduction reactions $\gamma^* N \to N^* \to M + B$ provide a tool to probe the inner structure 
of the contributing $N^*$ resonances through the extraction of the amplitudes for the transition 
between the virtual photon-nucleon initial state and the excited $N^*$ state, i.e. the $\gamma_vNN^*$ 
electrocoupling amplitudes, which are directly related to the $N^*$ structure. These electrocouplings 
can be represented by the so-called helicity amplitudes~\cite{helform}, among which are $A_{1/2}(Q^2)$ 
and $A_{3/2}(Q^2)$, which describe the $N^*$ resonance electroexcitation for the two different helicity 
configurations of a transverse photon and the nucleon, as well as $S_{1/2}(Q^2)$, which describes the 
$N^*$ resonance electroexcitation by longitudinal photons of zero helicity. Detailed comparisons of the 
theoretical predictions for these amplitudes with their experimental measurements form the basis of 
progress toward gauging our understanding of non-perturbative QCD. The measurement of the $\gamma_vNN^*$ 
electrocouplings is needed in order to gain access to the dynamical momentum-dependent mass and structure 
of the dressed quark in the non-perturbative domain where the quark-gluon coupling is large~\cite{vm1},
through mapping of the dressed quark mass function~\cite{vm2} and extractions of the quark distribution 
amplitudes for $N^*$ states of different quantum numbers~\cite{vm3}. This is critical in exploring the 
nature of quark-gluon confinement and dynamical chiral symmetry breaking (DCSB) in baryons. 

\begin{figure}[htbp]
\vspace{4.0cm}
\includegraphics{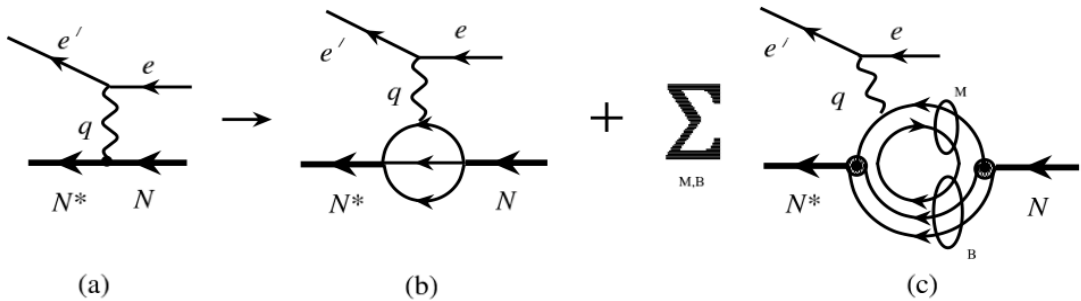}
\caption{Schematic representation of the $\gamma^*N \to N^*$ electroproduction process. (a) The fully 
dressed $\gamma_vNN^*$ electrocoupling that determines the $N^*$ contribution to the resonant part of 
the meson electroproduction amplitude. (b) The contribution of the three-quark core. (c) The 
contribution from the meson-baryon cloud, where the sum is over all intermediate meson and baryon 
states. This figure is taken from Ref.~\cite{review}.}
\label{electrocoupling} 
\end{figure}

Figure~\ref{electrocoupling} illustrates the two contributions to the $\gamma_vNN^*$ electrocouplings. 
In Fig.~\ref{electrocoupling}(b) the virtual photon interacts directly with the constituent quark, an 
interaction that is sensitive to the quark current and depends on the quark-mass function. However, the 
full meson electroproduction amplitude in Fig.~\ref{electrocoupling}(a) requires contributions to the 
$\gamma_vNN^*$ vertex from both non-resonant meson electroproduction and the hadronic scattering 
amplitudes as shown in Fig.~\ref{electrocoupling}(c). These contributions incorporate all possible 
intermediate meson-baryon states and all possible meson-baryon scattering processes that eventually 
result in the $N^*$ formation in the intermediate state of the reaction. These two contributions can be 
separated from each another using, for example, a coupled-channel reaction model~\cite{kamano}.

Current theoretical approaches to understand $N^*$ structure fall into two broad categories. In the 
first category are those that enable direct connection to the QCD Lagrangian, such as Lattice QCD (LQCD) 
and QCD applications of the Dyson-Schwinger equations (DSE). In the second category are those that use 
models inspired by or derived from our knowledge of QCD, such as quark-hadron duality, light-front 
holographic QCD (AdS/QCD), light-cone sum rules (LCSR), and CQMs. See Ref.~\cite{review} for an overview 
of these different approaches. It is important to realize that even those approaches that attempt to 
solve QCD directly can only do so approximately, and these approximations ultimately represent limitations 
that need careful consideration. As such, it is imperative that whenever possible the results of these 
intensive and challenging calculations be compared directly to the data on resonance electrocouplings from 
electroproduction experiments over a broad range of $Q^2$ for $N^*$ states with different quantum numbers.

\boldmath
\section{CLAS $N^*$ Program}
\unboldmath

Studies of the structure of the excited nucleon states, the so-called $N^*$ program, is one of the key
cornerstones of the physics program in Hall~B at Jefferson Laboratory (JLab). The large acceptance 
spectrometer CLAS~\cite{mecking}, which began data taking in 1997 and was decommissioned in 2012, was 
designed to measure photo- and electroproduction cross sections and polarization observables for beam
energies up to 6~GeV over a broad kinematic range for a host of different exclusive reaction channels. 
Consistent determination of $N^*$ properties from different exclusive channels with different couplings 
and non-resonant backgrounds offers model-independent support for the findings.

To date photoproduction data sets from CLAS and elsewhere have been used extensively to constrain
coupled-channel fits and advanced single-channel models. However, data at $Q^2$=0 allows us to identify 
$N^*$ states and determine their quantum numbers, but tell us very little about the structure of these 
states. It is the $Q^2$ dependence of the $\gamma_vNN^*$ electrocouplings that unravel and reveal these 
details. In addition, electrocoupling data are promising for studies of nucleon excited states as the ratio 
of resonant to non-resonant amplitudes increases with increasing $Q^2$. Finally, the electroproduction 
data are an effective tool to confirm the existence of new $N^*$ states as the data must be described by 
$Q^2$-independent resonance masses and hadronic decay widths.

The goal of the $N^*$ program with CLAS is to study the spectrum of $N^*$ states and their associated 
structure over a broad range of distance scales through studies of the $Q^2$ dependence of the 
$\gamma_vNN^*$ electrocouplings. For each final state this goal is realized through two distinct phases. 
The first phase consists of the measurements of the cross sections and polarization observables in as fine 
a binning in the relevant kinematic variables $Q^2$, $W$, $d\tau_{hadrons}$ (where $d\tau_{hadrons}$ 
represents the phase space of the final state hadrons) as the data support. The second phase consists of 
developing advanced reaction models that completely describe the data over its full phase space in order 
to then extract the electrocoupling amplitudes for the dominant contributing $N^*$ states. 

\begin{figure}[htbp]
\vspace{5.0cm}
\includegraphics{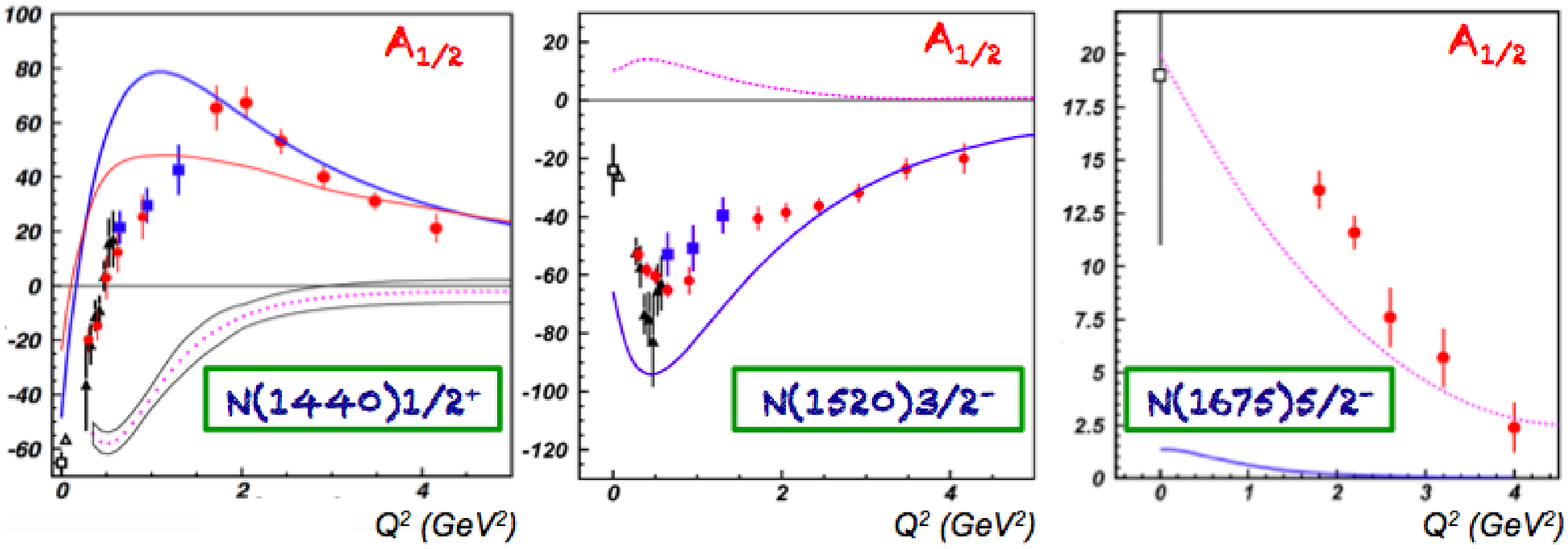}
\caption{The $A_{1/2}$ electrocoupling amplitudes (in units of 10$^{-3}$~GeV$^{-1/2}$) vs. $Q^2$ (GeV$^2$)
for the $N^*$ states $N(1440)\frac{1}{2}^+$ (left), $N(1520)\frac{3}{2}^-$ (middle), and 
$N(1675)\frac{5}{2}^-$ (right) from analyses of the CLAS $\pi N$ (circles) and $\pi\pi N$ (triangles, 
squares) data. (Left) Calculation from a non-relativistic light-front quark model with a running quark 
mass (red line) and calculation of the quark core from the DSE approach (blue line). (Middle/Right) 
Calculations from the hypercentral constituent quark model (blue lines). The magnitude of the 
meson-baryon cloud contributions is shown by the magenta line (or band) on each plot. See 
Refs.~\cite{review,mokeev13,mokeev15,ect15} for details on the data and the models.}
\label{low-lying} 
\end{figure}

Electrocoupling amplitudes for most $N^*$ states below 1.8~GeV have been extracted for the first time 
from analysis of CLAS data in the exclusive $\pi^+ n$ and $\pi^0 p$ channels for $Q^2$ up to 5~GeV$^2$, 
in $\eta p$ for $Q^2$ up to 4~GeV$^2$, and for $\pi^+ \pi^- p$ for $Q^2$ up to 1.5~GeV$^2$. 
Figure~\ref{low-lying} shows representative CLAS data for the $A_{1/2}$ electrocouplings for the
$N(1440)\frac{1}{2}^+$, $N(1520)\frac{3}{2}^-$, and $N(1675)\frac{5}{2}^-$
\cite{review,mokeev13,mokeev15,ect15}. Studies of the electrocouplings for $N^*$ states of different 
quantum numbers at lower $Q^2$ have revealed a very different interplay between the inner quark core 
and the meson-baryon cloud as a function of $Q^2$. Structure studies of the low-lying $N^*$ states, 
e.g. $\Delta(1232)\frac{3}{2}^+$, $N(1440)\frac{1}{2}^+$, $N(1520)\frac{3}{2}^-$, and $N(1535)\frac{1}{2}^-$, 
have made significant progress in recent years due to the agreement of results from independent analyses 
of the CLAS $\pi N$ and $\pi\pi N$ final states~\cite{mokeev13}. The good agreement of the extracted 
electrocouplings from both the $\pi N$ and $\pi \pi N$ exclusive channels is non-trivial in that these 
channels have very different mechanisms for the non-resonant backgrounds. The agreement thus provides 
compelling evidence for the reliability of the results.

The size of the meson-baryon dressing amplitudes are maximal for $Q^2 < 1$~GeV$^2$ (see 
Fig.~\ref{low-lying}). For increasing $Q^2$, there is a gradual transition to the domain where the quark 
degrees of freedom begin to dominate, as seen by the improved description of the $N^*$ electrocouplings 
obtained within the DSE approach, which accounts only for the quark core contributions. For 
$Q^2 > 5$~GeV$^2$, the quark degrees of freedom are expected to fully dominate the $N^*$ states
\cite{review}. Therefore, in the $\gamma_vNN^*$ electrocoupling studies for $Q^2 > 5$~GeV$^2$ expected 
with the future CLAS12 program (see Section~\ref{clas12-program}), the quark degrees of freedom will be 
probed more directly with only small contributions from the meson-baryon cloud.

\begin{figure}[htbp]
\vspace{4.7cm}
\includegraphics{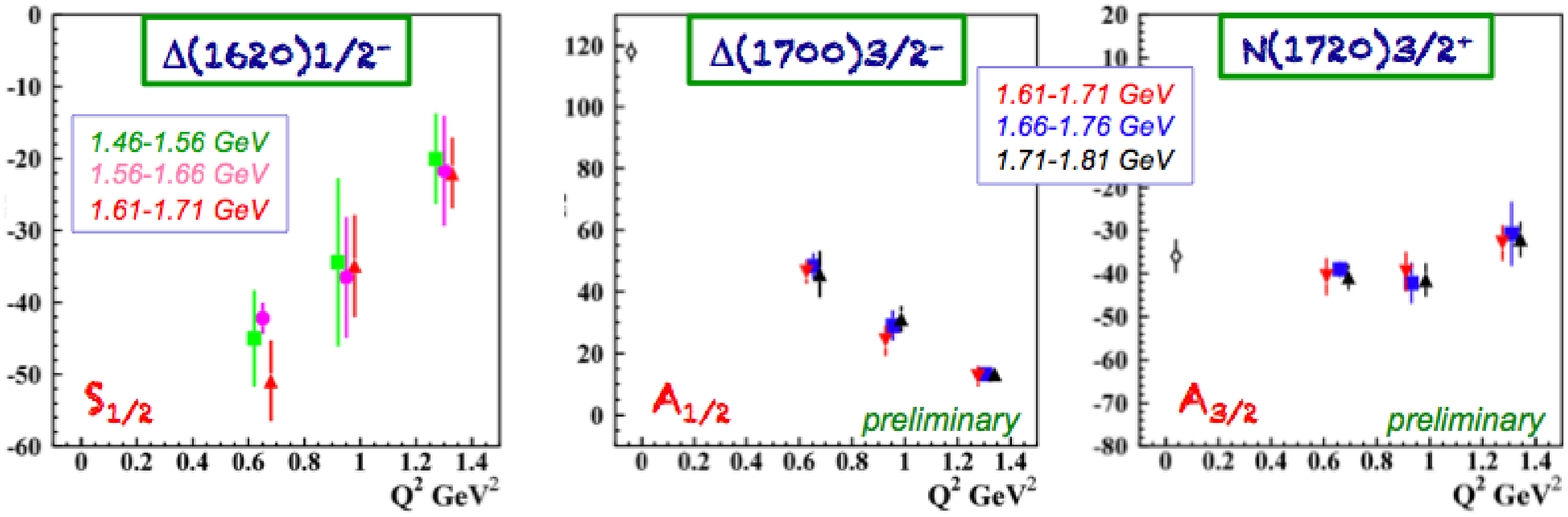}
\caption{CLAS results for the $N^*$ electrocoupling amplitudes (in units of 10$^{-3}$~GeV$^{-1/2}$) from 
analysis of the exclusive $\pi^+\pi^- p$ final state as a function of $Q^2$ (GeV$^2$). (Left) $S_{1/2}$ 
of the $\Delta(1620)\frac{1}{2}^-$~\cite{mokeev15}, (middle) preliminary extraction of $A_{1/2}$ for the 
$\Delta(1700)\frac{3}{2}^-$~\cite{ect15}, and (right) preliminary extraction of $A_{3/2}$ for the 
$N(1720)\frac{3}{2}^+$~\cite{ect15}. Each electrocoupling amplitude was extracted in independent fits in 
different bins of $W$ across the resonance peak width as shown for each $Q^2$ bin (points in each $Q^2$ 
bin offset for clarity).}
\label{high-lying} 
\end{figure}

Analysis of CLAS data for the $\pi \pi N$ channel has provided the only detailed structural information 
available regarding higher-lying $N^*$ states, e.g. $\Delta(1620)\frac{1}{2}^-$, $N(1650)\frac{1}{2}^-$, 
$N(1680)\frac{5}{2}^+$, $\Delta(1700)\frac{3}{2}^-$, and $N(1720)\frac{3}{2}^+$. Fig.~\ref{high-lying} 
shows a representative set of illustrative examples for $S_{1/2}$ for the $\Delta(1620)\frac{1}{2}^-$
\cite{mokeev15}, as well as for $A_{1/2}$ for the $\Delta(1700)\frac{3}{2}^-$ and $A_{3/2}$ for the 
$N(1720)\frac{3}{2}^+$~\cite{ect15}. Here the analysis for each $N^*$ state was carried out independently 
in different bins of $W$ across the width of the resonance for $Q^2$ up to 1.5~GeV$^2$ with very good 
correspondence within each $Q^2$ bin. Note that most of the $N^*$ states with masses above 1.6~GeV decay 
preferentially through the $\pi \pi N$ channel instead of the $\pi N$ channel.

\begin{figure}[htbp]
\vspace{8.3cm}
\includegraphics{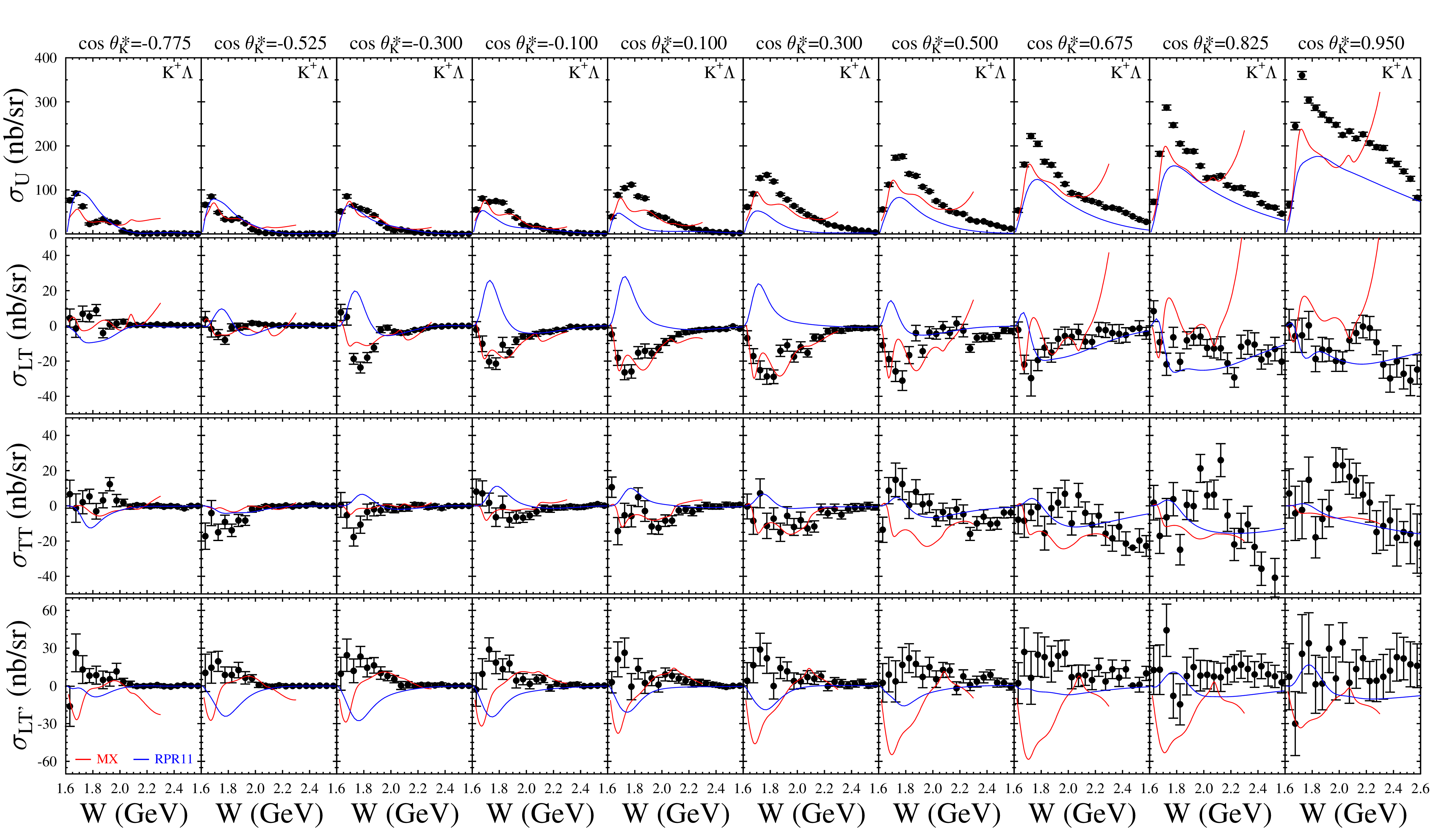}
\caption{Structure functions $\sigma_U = \sigma_T + \epsilon \sigma_L$, $\sigma_{LT}$, $\sigma_{TT}$, 
and $\sigma_{LT'}$ (nb/sr) for $K^+\Lambda$ production vs. $W$ (GeV) for $E_{beam}$=5.5~GeV for 
$Q^2$=1.80~GeV$^2$ and $\cos \theta_K^*$ values as shown from CLAS data~\cite{carman3,clasdb}. The error 
bars represent the statistical uncertainties only. The red curves are from the hadrodynamic $KY$ model 
of Maxwell~\cite{max12a} and the blue curves are from the hybrid RPR-2011 $KY$ model from Ghent
\cite{decruz}.}
\label{lam_q1_w} 
\end{figure}

\begin{figure}[htbp]
\vspace{8.3cm}
\includegraphics{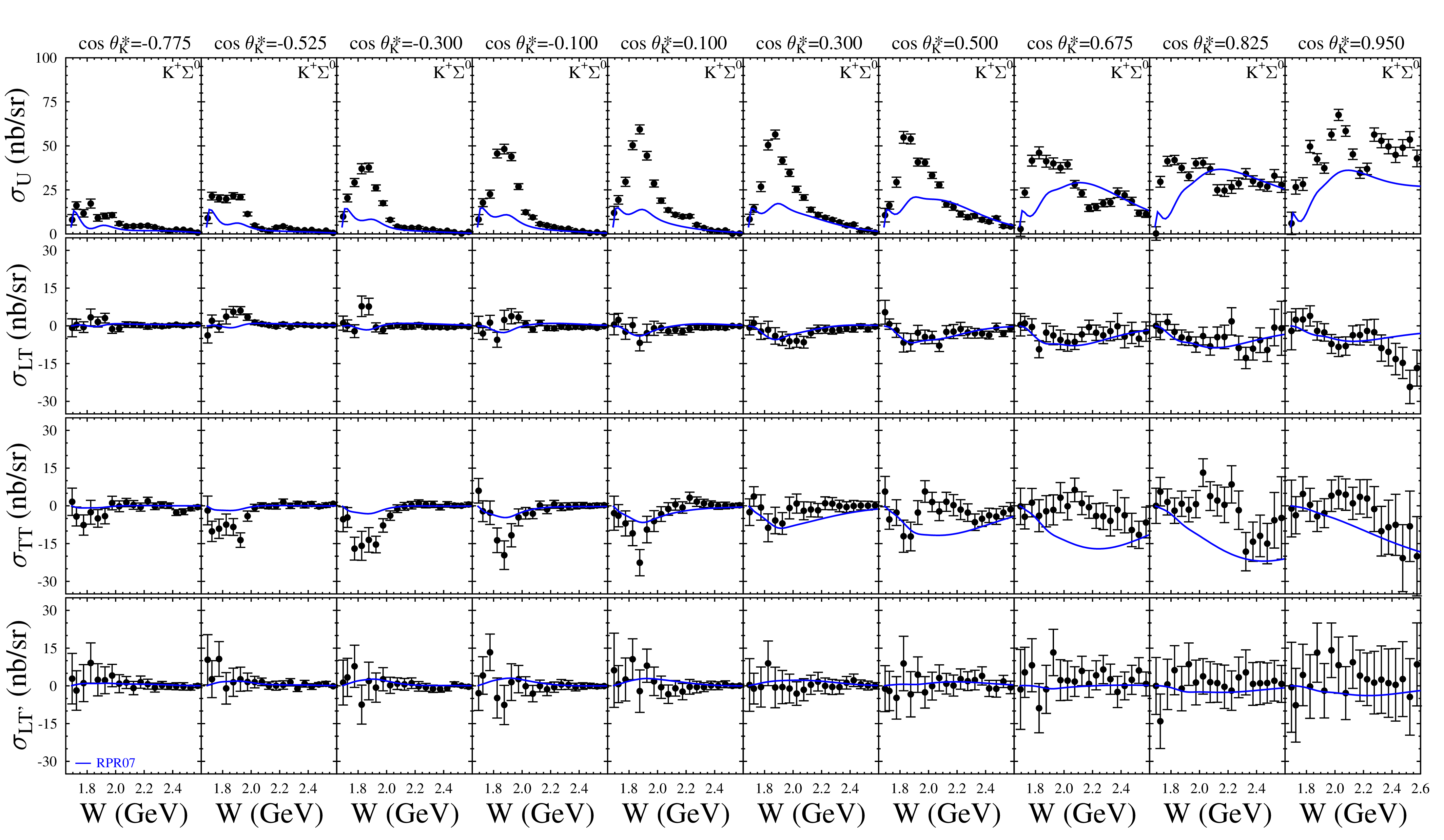}
\caption{Structure functions $\sigma_U = \sigma_T + \epsilon \sigma_L$, $\sigma_{LT}$, $\sigma_{TT}$, 
and $\sigma_{LT'}$ (nb/sr) for $K^+\Sigma^0$ production vs. $W$ (GeV) for $E_{beam}$=5.5~GeV for 
$Q^2$=1.80~GeV$^2$ and $\cos \theta_K^*$ values as shown from CLAS data~\cite{carman3,clasdb}. The error 
bars represent the statistical uncertainties only. The blue curves are from the hybrid RPR-2007 $KY$ model 
from Ghent~\cite{corthals}.}
\label{sig_q1_w} 
\end{figure}

With a goal to have an independent determination of the electrocouplings for each $N^*$ state from
multiple exclusive reaction channels, a natural avenue to investigate for the higher-lying $N^*$ states is 
the strangeness channels $K^+\Lambda$ and $K^+\Sigma^0$. In fact, data from the $KY$ channels are critical 
to provide an independent extraction of the electrocoupling amplitudes for the higher-lying $N^*$ states. 
The CLAS program has yielded by far the most extensive and precise measurements of $KY$ electroproduction 
data ever measured across the nucleon resonance region. These measurements have included the separated 
structure functions $\sigma_T$, $\sigma_L$, $\sigma_U = \sigma_T + \epsilon \sigma_L$, $\sigma_{LT}$, 
$\sigma_{TT}$, and $\sigma_{LT'}$ for $K^+\Lambda$ and $K^+\Sigma^0$~\cite{raue-car,5st,sltp,carman3}, 
recoil polarization for $K^+\Lambda$~\cite{ipol}, and beam-recoil transferred polarization for $K^+\Lambda$ 
and $K^+\Sigma^0$~\cite{carman1,carman2}. For the hyperon polarization measurements, we have taken
advantage of the self-analyzing nature of the weak decay of the $\Lambda$. These measurements span $Q^2$ 
from 0.5 to 4.5~GeV$^2$, $W$ from 1.6 to 3.0~GeV, and the full center-of-mass angular range of the $K^+$. 
The $KY$ final states, due to the creation of an $s\bar{s}$ quark pair in the intermediate state, are 
naturally sensitive to coupling to higher-lying $s$-channel resonance states at $W > 1.6$~GeV, a region 
where our knowledge of the $N^*$ spectrum is the most limited. Note also that although the two ground-state 
hyperons have the same valence quark structure ($uds$), they differ in isospin, such that intermediate 
$N^*$ resonances can decay strongly to $K^+\Lambda$ final states, but intermediate $\Delta^*$ states 
cannot. Because $K^+\Sigma^0$ final states can have contributions from both $N^*$ and $\Delta^*$ states, 
the hyperon final state selection constitutes an isospin filter. Shown in Figs.~\ref{lam_q1_w} and 
\ref{sig_q1_w} is a small sample of the available data in the form of the $K^+\Lambda$ and $K^+\Sigma^0$ 
structure functions $\sigma_U$, $\sigma_{LT}$, $\sigma_{TT}$, and $\sigma_{LT'}$~\cite{carman3,clasdb}, 
illustrating its broad kinematic coverage and statistical precision.

While there has been progress toward a better understanding of the low-lying $N^*$ states in the region 
below 1.6~GeV, the vast majority of the predicted missing $N^*$ and $\Delta^*$ states lie in the region 
from $1.6 < W < 3$~GeV. To date the PDG lists only four $N^*$ states, $N(1650)\frac{1}{2}^-$, 
$N(1710)\frac{1}{2}^+$, $N(1720)\frac{3}{2}^+$, and $N(1900)\frac{3}{2}^+$, with known couplings to 
$K\Lambda$ and no $N^*$ states are listed that couple to $K\Sigma$~\cite{pdg}; only a single $\Delta^*$ 
state, $\Delta(1920)\frac{3}{2}^+$, is listed with coupling strength to $K\Sigma$. The branching ratios 
to $KY$ provided for these states are typically less than 10\% with uncertainties on the order of the 
measured coupling. While the relevance of this core set of $N^*$ states in the $\gamma^{(*)} p \to K^+ \Lambda$ 
reaction has long been considered a well-established fact, this set of states falls well short of reproducing 
the experimental results for $W < 2$~GeV. 

Figs.~\ref{lam_q1_w} and \ref{sig_q1_w} include two of the more advanced single channel reaction models 
for the electromagnetic production of $KY$ final states. The MX model is the isobar model from Maxwell
\cite{max12a}, and the RPR-2007~\cite{corthals} and RPR-2011~\cite{decruz} models are from the Ghent Regge 
plus Resonance (RPR) framework. Both the MX and RPR models were developed based on fits to the extensive and 
precise photoproduction data from CLAS and elsewhere and describe those data reasonably well. However, they 
utterly fail to describe the electroproduction data in any of the kinematic phase space. Reliable information 
on $KY$ hadronic decays from $N^*$s is not yet available due to the lack of an adequate reaction model. 
However, after such a model is developed, the $N^*$ electrocoupling amplitudes for states that couple to 
$KY$ can be obtained from fits to the extensive existing CLAS $KY$ electroproduction data over the range 
$0.5 < Q^2 < 4.5$~GeV$^2$, which should be carried out independently in different bins of $Q^2$ with the
same $KY$ hadronic decays, extending the available information on these $N^*$ states. The development 
of reaction models for the extraction of the $\gamma_vNN^*$ electrocouplings from the $KY$ electroproduction 
channels is urgently needed.

It is also important to note that the $\pi N$ and $\pi\pi N$ electroproduction channels represent the two 
dominant exclusive channels in the resonance region. The knowledge of the electroproduction mechanisms for 
these channels is critically important for $N^*$ studies in channels with smaller cross sections such as 
$K^+\Lambda$ and $K^+\Sigma^0$ production, as they can be significantly affected in leading order by 
coupled-channel effects produced by their hadronic interactions in the pionic channels. Ultimately such
effects need to be properly included in the $KY$ reaction models.

\boldmath
\section{CLAS12 $N^*$ Program}
\label{clas12-program}
\unboldmath

As part of the upgrade of the JLab accelerator from a maximum electron beam energy of 6~GeV to a
maximum energy of 12~GeV, a new large acceptance spectrometer called CLAS12 was designed for experimental
Hall~B to replace the CLAS spectrometer. The new CLAS12 spectrometer~\cite{clas12} is designed for
operation at beam energies up to 11~GeV (the maximum possible for delivery to Hall~B) and will operate at
a nominal beam-target luminosity of $1\times10^{35}$~cm$^{-2}$s$^{-1}$, an order of magnitude increase 
over previous CLAS operation. This luminosity will allow for precision measurements of cross sections and 
polarization observables for many exclusive reaction channels for invariant energy $W$ up to 3~GeV,
the full decay product phase space, and four-momentum transfer $Q^2$ up to 12~GeV$^2$. The physics program 
for CLAS12 has focuses on measurements of the spatial and angular momentum structure of the nucleon, 
investigation of quark confinement and hadron excitations, and studies of the strong interaction in nuclei. 
The commissioning of the new CLAS12 spectrometer is scheduled to take place in the first part of 2017, 
followed shortly thereafter by the first physics running period.
 
The electrocoupling parameters determined for several low-lying $N^*$ states from the data involving the 
pionic channels for $Q^2$ up to 5~GeV$^2$ have already provided valuable information. At these distance 
scales, the resonance structure is determined by both meson-baryon dressing and dressed quark contributions. 
The $N^*$ program with the new CLAS12 spectrometer in Hall~B is designed to study excited nucleon structure 
over a broad range of $Q^2$, from $Q^2$ = 3~GeV$^2$ to allow for direct overlap with the data sets collected 
with the CLAS spectrometer, up to $Q^2$=12~GeV$^2$, the highest photon virtualities ever probed in exclusive 
electroproduction reactions. In the kinematic domain of $Q^2$ from 3 to 12~GeV$^2$, the data can probe more
directly the inner quark core and map out the transition from the confinement to the perturbative QCD domains.

The $N^*$ program with CLAS12 consists of two approved experiments. E12-09-003~\cite{e12-09-003} will 
focus on the non-strange final states (primarily $\pi N$, $\eta N$, $\pi \pi N$) and E12-06-108A
\cite{e12-06-108a} will focus on the strange final states (primarily $K^+\Lambda$ and $K^+\Sigma^0$). 
These experiments will allow for the determination of the $Q^2$ evolution of the electrocoupling parameters 
for $N^*$ states with masses in the range up to 3~GeV in the regime of $Q^2$ up to 12~GeV$^2$. These 
experiments will be part of the first production physics running period with CLAS12 in 2017. The experiments 
will collect data simultaneously using a longitudinally polarized 11~GeV electron beam on an unpolarized 
liquid-hydrogen target.

The program of $N^*$ studies with the CLAS12 detector has a number of important objectives. These include:

\vskip 0.3cm

i) To map out the quark structure of the dominant $N^*$ and $\Delta^*$ states from the acquired
electroproduction data through the exclusive final states including the non-strange channels $\pi^0 p$, 
$\pi^+ n$, $\eta p$, $\pi^+\pi^- p$, as well as the dominant strangeness channels $K^+\Lambda$ and 
$K^+\Sigma^0$. This objective is motivated by results from existing analyses such as those shown in 
Fig.~\ref{low-lying}, where it is seen that the meson-baryon dressing contribution to the $N^*$ structure 
decreases rapidly with increasing $Q^2$. The data can be described approximately in terms of dressed quarks 
already for $Q^2$ up to 3~GeV$^2$. It is therefore expected that the data at $Q^2 > 5$~GeV$^2$ can be 
used more directly to probe the quark substructure of the $N^*$ and $\Delta^*$ states~\cite{review}. The 
comparison of the extracted resonance electrocoupling parameters from this new higher $Q^2$ regime to the 
predictions from LQCD and DSE calculations will allow for a much improved understanding of how the internal 
dressed quark core emerges from QCD and how the dynamics of the strong interaction are responsible for the 
formation of $N^*$ and $\Delta^*$ states of different quantum numbers.

\vskip 0.15cm

ii) To investigate the dynamics of dressed quark interactions and how they emerge from QCD to generate
$N^*$ states of different quantum numbers. This work is motivated by recent advances in the DSE approach
\cite{bhagwat,roberts} and LQCD~\cite{bowman}, which have provided links between the dressed quark 
propagator, the dressed quark scattering amplitudes, and the QCD Lagrangian. These approaches also relate
the momentum dependence of the dressed quark mass function to the $\gamma_vNN^*$ electrocouplings for $N^*$
states of different quantum numbers. DSE analyses of the extracted $N^*$ electrocoupling parameters have 
the potential to allow for investigation of the origin of quark-gluon confinement in baryons and the 
nature of more than 98\% of the hadron mass generated non-perturbatively through DSCB, since both of these 
phenomena are rigorously incorporated into the DSE approach~\cite{review}. Efforts are currently underway to 
study the sensitivity of the proposed electromagnetic amplitude measurements to different parameterizations 
of the momentum dependence of the quark mass~\cite{cloet}.

\vskip 0.15cm

iii) To offer constraints from resonance electrocoupling amplitudes on the Generalized Parton Distributions
(GPDs) describing $N \to N^*$ transitions. We note that a key aspect of the CLAS12 measurement program is 
the characterization of exclusive reactions at high $Q^2$ in terms of GPDs. The elastic and $\gamma_vNN^*$ 
transition form factors represent the first moments of the GPDs~\cite{frankfurt,goeke}, and they provide 
for unique constraints on the structure of nucleons and their excited states. Thus the $N^*$ program at 
high $Q^2$ represents the initial step in a reliable parameterization of the transition $N \to N^*$ GPDs 
and is an important part of the larger overall CLAS12 program studying exclusive reactions.

\section{Concluding Remarks}

The study of the spectrum and structure of the excited nucleon states represents one of the key physics 
foundations for the measurement program in Hall~B with the CLAS spectrometer. To date measurements
with CLAS have provided a dominant amount of precision data (cross sections and polarization observables)
for a number of different exclusive final states for $Q^2$ from 0 to 4.5~GeV$^2$. From the $\pi N$
and $\pi \pi N$ data, the electrocouplings of most $N^*$ states up to $\sim$1.8~GeV have been extracted 
for the first time. With the development and refinement of reaction models to describe the extensive CLAS
$K^+\Lambda$ and $K^+\Sigma^0$ electroproduction data, the data from the strangeness channels is expected
to provide an important complement to study the electrocoupling parameters for higher-lying $N^*$
resonances with masses above 1.6~GeV.

The $N^*$ program with the new CLAS12 spectrometer will extend these studies up to $Q^2$ of 12~GeV$^2$,
the highest photon virtualities ever probed in exclusive reactions. This program will ultimately focus
on the extraction of the $\gamma_vNN^*$ electrocoupling amplitudes for the $s$-channel resonances that 
couple strongly to the non-strange final states $\pi N$, $\eta N$, and $\pi \pi N$, as well as the strange
$K^+\Lambda$ and $K^+\Sigma^0$ final states. These studies in concert with theoretical developments will 
allow for insight into the strong interaction dynamics of dressed quarks and their confinement in baryons 
over a broad $Q^2$ range. The data will address the most challenging and open problems of the Standard Model 
on the nature of hadron mass, quark-gluon confinement, and the emergence of the $N^*$ states of different 
quantum numbers from QCD.

\vskip 0.5cm

{\bf Acknowledgments:}  This work was supported by the U.S. Department of Energy. The author is grateful 
for many lengthy and fruitful discussions on this topic with Victor Mokeev and Ralf Gothe. The author also 
thanks the organizers of the ECT* 2015 Workshop Nucleon Resonances: From Photoproduction to High Photon 
Virtualities for the opportunity to present this work and participate in this workshop.

\vskip 0.5cm


\end{document}